\documentclass[11pt]{article}

\usepackage{cite}
\usepackage{color,fontenc,float}
\usepackage{tikz}
\usepackage[normalem]{ulem}
\usepackage{amsmath,amsthm,amssymb,framed}
\usepackage{mathtools,hyperref}
\newcommand{\rem}[1]{}
\newcommand{\bom}{\mbox{\boldmath$\omega$}}
\newcommand{\bu}{\mbox{\boldmath$u$}}

\newcommand{\br}{\mbox{\boldmath$r$}}
\newcommand{\bx}{\mbox{\boldmath$x$}}
\newcommand{\bdf}{\mbox{\boldmath$f$}}

\newcommand{\Dim}{\mathfrak{D}_{m}}
\newcommand{\Cim}{\mathfrak{C}_{m}}
\newcommand{\Sp}{S_{\!p}}

\newcommand{\bel}{\begin{equation}\label}
\newcommand{\ee}{\end{equation}}
\newcommand{\beq}{\begin{eqnarray}\label} 
\newcommand{\eeq}{\end{eqnarray}} 
\newcommand{\bc}{\begin{center}} 
\newcommand{\ec}{\end{center}} 
\newcommand{\ben}{\begin{enumerate}}
\newcommand{\een}{\end{enumerate}}
\newcommand{\bit}{\begin{itemize}}
\newcommand{\eit}{\end{itemize}}

\newcommand\third{\ensuremath{{\scriptstyle\frac{1}{3}}}}
\newcommand\twothirds{\ensuremath{{\scriptstyle\frac{2}{3}}}}

\newcommand\fourfifths{\ensuremath{{\scriptstyle\frac{4}{5}}}}

\theoremstyle{definition}

\numberwithin{equation}{section}
\rem{
\setcounter{section}{0}
\setcounter{subsection}{0}
\setcounter{theorem}{1}
\setcounter{lemma}{1}
\setcounter{corollary}{1}
\setcounter{proposition}{0}
\setcounter{page}{1}
\makeatletter
\@addtoreset{equation}{section}

\makeatother
}
\begin{document}

\bc
\textbf{\color{blue}Identifying the multifractal set on which energy dissipates\\
in a turbulent Navier-Stokes fluid}
\par\vspace{3mm}
\textbf{John D. Gibbon}
\par\vspace{3mm}
Department of Mathematics, Imperial College London, London SW7 2AZ, UK
\ec


\begin{abstract}
The rich multifractal properties of fluid turbulence illustrated by the work of Parisi and Frisch are related explicitly to Leray's weak solutions of the three-dimensional Navier-Stokes equations. Directly from this correspondence it is found that the set on which energy dissipates, $\mathbb{F}_{m}$, has a range of dimensions $\Dim=3/m$ ($1 \leq m \leq \infty$), and a corresponding range of sub-Kolmogorov dissipation inverse length scales $L\eta_{m}^{-1} \leq Re^{3/(1+\Dim)}$ spanning $Re^{3/4}$ to $Re^{3}$. Correspondingly, the multifractal model scaling parameter $h$, must obey $h \geq h_{min}$ with $-\twothirds \leq h_{min} \leq \third$.
\end{abstract}
\bc
\textit{In memory of Charlie Doering (1956-2021)}
\ec

\section{\large The need for a mathematical telescope}\label{Intro}

Despite the clich\'e that \textit{``turbulence is the last great unsolved problem in classical physics''} \cite{SF2006}, the challenge that it poses to physicists, mathematicians and engineers remains 
\cite{PF1985,Frisch1995,MS1991,Verma2019,BD2019,PKYR2020,EIH2020,BPBY2019,IMYUK2016,BBBLR2016,DGGKPV2014,SM1986,
DCB1991,FrischVerg1992,DF2002,AD2006,CRLMPA2005,IGK2009}. A natural correspondence is proposed that helps us relate the rich multifractality of the statistical properties of homogeneous and isotropic fluid turbulence~\cite{PF1985,Frisch1995,MS1991} to the methods of mathematical analysis which are used to study weak solutions of the incompressible Navier-Stokes equations  \cite{Leray1934,FGT1981,DG1995,ESS2003,FMRT2001,CS2014,RRS2016,JDG2019,JDG2020}, including the Millenial regularity problem \cite{NS-Clay}. Heretofore the two approaches have taken divergent paths which has caused a lacuna in our understanding. Multifractality is manifest in the wide sweep of vortical structures that appear in turbulent flows, ranging from three-dimensional vortices at many scales, to semi-broken filaments whose fractal dimension is less than unity\,: see Fig. 1 for illustrative examples. In telescopic terms, the simultaneous existence of many vortical structures each with its own fractal dimension requires the mathematical equivalent of an adjustable focus, somewhat in the manner of a \textit{zoom lens}. Thus, an adjustable parameter with a wide range of values is needed. The multifractal model (MFM) of Parisi and Frisch possesses this latter feature through the allowed variation of its scaling parameter $h$ \cite{PF1985}, while the Navier-Stokes equations seemingly do not. The MFM was developed after it had become clear that Kolmogorov's 1941 theory needed to be modified in order to take account of the effects of intermittency \cite{Frisch1995}\,: see also \cite{Benzi1984,Kraichnan1985,PV1987a,PV1987b,Nelkin1990,Eyink2008}. 
\par\smallskip
\begin{figure}[ht]
\centering
\includegraphics[scale=0.10]{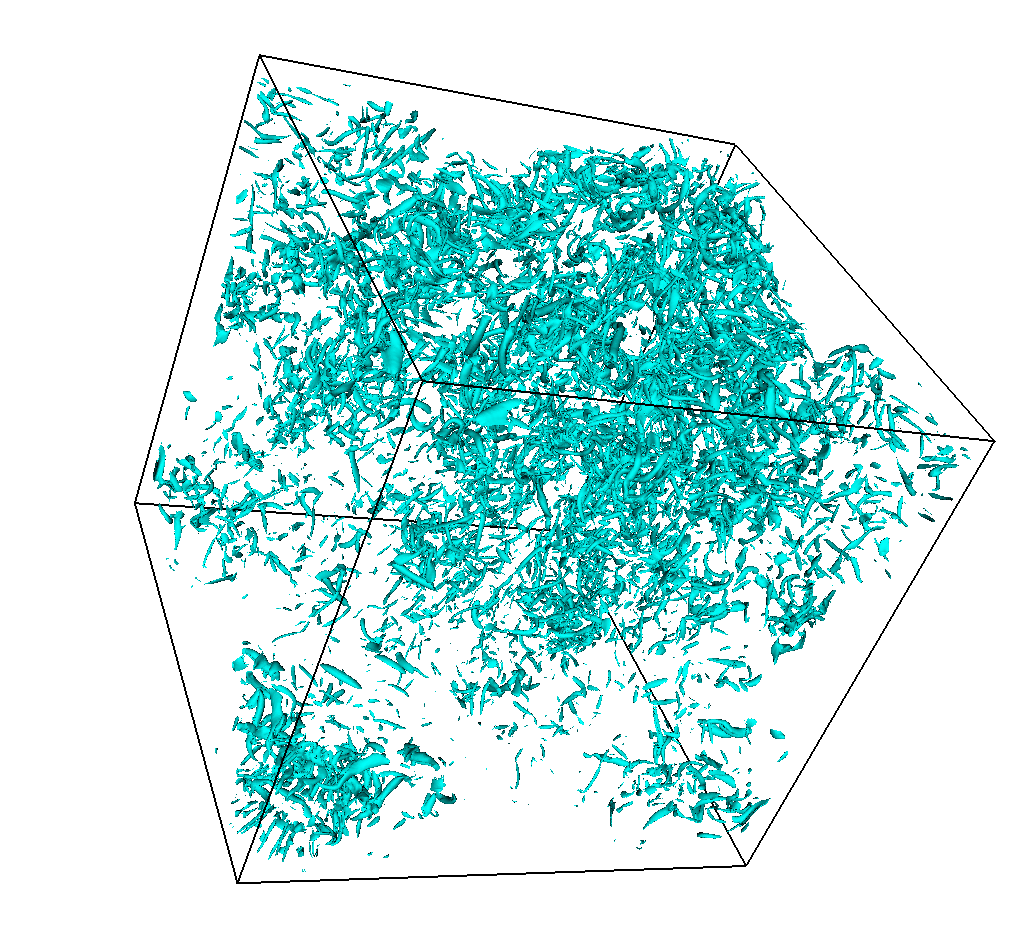}\quad 
\includegraphics[scale=0.30]{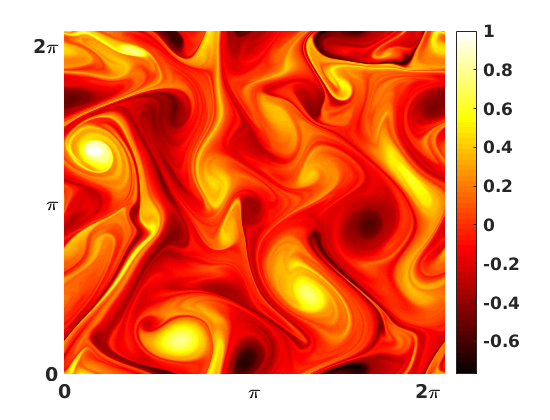}\quad
\includegraphics[scale=0.30]{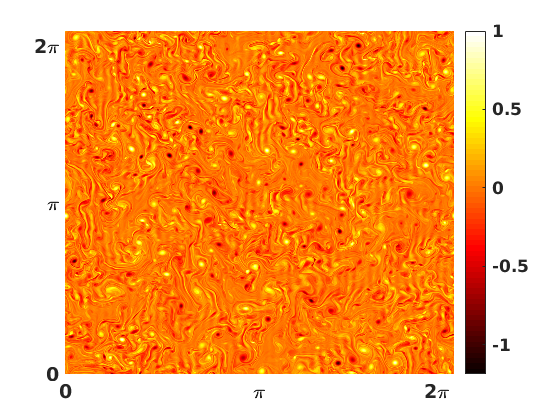}
\vspace{0mm}
\caption{\scriptsize (left) Iso-surface plot of the magnitude of the vorticity $|\bom|$ (two standard deviations above the mean) from a $512^{3}$ pseudospectral direct numerical simulation of the unforced $3D$ Navier-Stokes equations at a representative time at an integral-scale Reynolds number of 874. Plot courtesy of Nadia B. Padhan, IISc, Bangalore. The second and third are filled-contour plots of the vorticity from $2048^{2}$ pseudospectral direct numerical simulations of the forced $2D$ Navier-Stokes equations in a statistically steady state, courtesy of Kiran Kolluru, IISc, Bangalore\,:  (middle) forward-cascade-dominated turbulence ($Re_{\lambda}=1957$)\,; (right) inverse-cascade-dominated turbulence ($Re_{\lambda}=1729$).}
\end{figure}
The question is this\,: can this telescopic property be derived from Leray's weak solution formulation of the three-dimensional incompressible Navier-Stokes equations \cite{Leray1934,FGT1981,DG1995,ESS2003,FMRT2001,CS2014,RRS2016,JDG2019,JDG2020}? From the Navier-Stokes side of the fence, no mathematical tools seemingly exist that would allow us to perform rigorous analysis on fractal sets. Standard methods of Navier-Stokes analysis involve integration over the full domain volume, which has the unfortunate consequence of washing out the delicate nature of the set on which energy dissipates \cite{Verma2019,BD2019}. This paper circumvents these problems by showing how the existence of a multifractal set on which energy dissipates is natural to the problem. Results can be extracted from the Leray formalism which are then compared to the multifractal theory of Parisi and Frisch \cite{PF1985}. 
\par\smallskip
In recent work, Dubrulle and Gibbon \cite{DG2022} addressed the issue of the equivalence of the Navier-Stokes equations and the multifractal model. They used the Paladin-Vulpiani\footnote{The argument used in \cite{PV1987a,PV1987b} to estimate this scale is to equate the turnover time to the viscous diffusion time. It is also based on a Reynolds number defined in terms of the energy dissipation rate.}  \cite{PV1987a,PV1987b} inverse dissipation length scale
\bel{PVscale}
L\eta^{-1}_{PV} \sim Re^{1/(1+h)}
\ee
as an \textit{ad hoc} bridge between the two theories, where $h$ is the multifractal scaling parameter. In this paper we are able to go a step further by showing that this scale, at least in an inequality form, appears naturally out of the correspondence between the two theories without any \textit{ad hoc} introduction.

\par\vspace{3mm}
\section{\large Recasting Leray's weak solution formulation of the Navier-Stokes equations} 

\subsection{A zoom lens}

The incompressible Navier-Stokes equations are 
\bel{NSE1}
\left(\partial_{t} + \bu\cdot\nabla\right)\bu + \nabla p = \nu \Delta\bu + \bdf(\bx)
\ee
subject to $\mbox{div}\,\bu = \mbox{div}\,\bdf= 0$, on a periodic domain $V_{d}=[0,\,L]^{d}$ for dimensions $d=3,\,2,\,1$. Results on weak solutions can be expressed in many ways \cite{Leray1934} but most appear in time averaged form  \cite{FGT1981,DG1995,ESS2003,FMRT2001,CS2014,RRS2016,JDG2019,JDG2020}. In \cite{JDG2019}, it was shown that for $1 \leq m \leq \infty$, the Sobolev norms 
\bel{Finv2}
\|\nabla^{n}\bu\|_{2m} = \left(\int_{V_{d}}|\nabla^{n}\bu|^{2m}\,dV\right)^{1/2m}\,,
\ee
under the Navier-Stokes invariance property $\bx' = \lambda^{-1}\bx$\,; $t' = \lambda^{-2}t$ and $\bu = \lambda^{-1}\bu'$, 
have the scaling property 
\beq{Finv3}
\|\nabla^{n}\bu\|_{2m} &=& \lambda^{-1/\alpha_{n,m,d}}\|\nabla^{'n}\bu'\|_{2m}\,\\
\alpha_{n,m,d} &=& \frac{2m}{2m(n+1)-d}\,.
\eeq
It has been shown in \cite{JDG2019,JDG2020} that for $n \geq 1$, the weighted time averages of the dimensionless family 
\bel{Fdefa}
F_{n,m,d} = \nu^{-1}L^{1/\alpha_{n,m,d}}\|\nabla^{n}\bu\|_{2m}
\ee
are finite such that\footnote{The proof of (\ref{F1a}) in \cite{JDG2019} depends heavily on the result in \cite{FGT1981}. The case $n=0$ is valid only for $m>3$.} 
\bel{F1a}
\left<F_{n,m,d}^{(4-d)\alpha_{n,m,d}}\right>_{T} < \infty\,.
\ee
The brackets $\left<\cdot\right>_{T}$ are a time average up to time $T>0$.  The set of exponents $\alpha_{n,m,d}$ in (\ref{F1a}) clearly play a significant role in Navier-Stokes regularity properties\footnote{It was shown in \cite{JDG2019} that to prove full regularity in the $d=3$ case a factor of $2\alpha_{n,m,3}$ would be needed in the exponent of (\ref{F1a}). This result remains elusive.}. A series of well known weak solution results can be recovered by choosing various values of $n$ and $m$, which all have the same level of regularity \cite{JDG2019}. Increasing $m$ amplifies the larger features in the solution in proportion to the smaller until $m=\infty$ is reached which corresponds to the maximum. Therefore, adjusting the value of $m$ acts as a focus control on a telescope that allows us to zoom into different features in a landscape. The subscript labelling on $\alpha_{n,m,d}$ encodes the nature of the norm of which it is an exponent\,: thus $(n,m,d)$ means \textit{``$n$ derivatives in $L^{2m}$ in integer-$d$ spatial dimensions''}. However, $\alpha_{n,m,d}$ can be rewritten as
\beq{F3}
\alpha_{n,m,d} &=& \frac{2}{2(n+1)-\Dim}\,,\qquad \Dim = d/m\,,\nonumber\\
&=& \alpha_{n,1,\Dim}\,.
\eeq
The subscripts on $\alpha_{n,1,\Dim}$ now suggest that the problem can now be recast into \textit{``$n$ derivatives in $L^{2}$ on a set, designated as $\mathbb{F}_{m}$, which has a range of dimensions $\Dim = d/m$".} The range of dimensions $\Dim = d/m$ suggests that $\mathbb{F}_{m}$ is multifractal in character, with a corresponding set of co-dimensions 
\bel{Cmdef}
\Cim = d\left(1-\frac{1}{m}\right)\,.
\ee
Mimicking (\ref{F1a}), the recasting of $\alpha_{n,m,d}$ into $\alpha_{n,1,\Dim}$ suggests that the family of time averages 
\bel{F4}
X_{n,\Dim}(T) = \left<F_{n,1,\Dim}^{(4-\Dim)\,\alpha_{n,1,\Dim}}\right>_{T}\,,
\ee
may be a key set of objects. The $X_{n,m,\Dim}$ are neither necessarily equal to the averages in (\ref{F1a}) nor would it be an easy task to prove they are bounded above because there exists no mechanism for performing analysis on fractal or multifractal domains. Nevertheless, the case $n=1$ is an exception in that it can be bounded because of the cancellation of the factor of $(4-\Dim)$ in 
\bel{F5}
\left[(4-\Dim)\alpha_{n,1,\Dim}\right]_{n=1} = 2\,,\qquad 1 \leq m \leq \infty\,,
\ee
for every value of $0 \leq \Dim \leq d$. It is this case we consider in the next subsection.

\subsection{The three-dimensional case when $n=1$}

The $n=1$ case of (\ref{F4}) for $d=3$ corresponds, for each $m$, to a time-averaged energy dissipation rate $\varepsilon_{m}$ on the multifractal set $\mathbb{F}_{m}$. Each $\varepsilon_{m}$ is defined as
\bel{F6}
\varepsilon_{m} = \nu L^{-\Dim}\left<\int_{\mathbb{F}_{m}}|\nabla\bu|^{2}dV\right>_{T}\,,\quad \Dim = 3/m\,.
\ee
Noting that $\mathbb{F}_{m} \subset \mathbb{T}^{3}$, (\ref{F6}) can be re-written as
\beq{F7a}
\varepsilon_{m} &\leq& \nu L^{3-\Dim}L^{-3}\left< \int_{\mathbb{T}^{3}}|\nabla\bu|^{2}dV\right>_{T}\nonumber\\
&\leq&  c\,L^{-1-\Dim}\nu^{3}Re^{3}\,.
\eeq
The $Re^{3}$ upper bound is derived for generic narrow-band forcing considered by Doering and Foias \cite{DF2002} where the Reynolds number $Re$ is defined by 
\bel{Redef}
Re = UL/\nu\qquad\mbox{with}\qquad U^{2} = L^{-3}\left<\|\bu\|^{2}_{2}\right>_{T}\,.
\ee\noindent
(\ref{F7a}) shows that $\varepsilon_{m}\nu^{-3}$ can be expressed in terms of a set of inverse lengths $\eta_{m}^{-1}$ defined such that 
\bel{F7b}
\varepsilon_{m}\nu^{-3} := \eta_{m}^{-1-\Dim}
\ee
leading to
\bel{F8}
L\eta_{m}^{-1} \leq c\,Re^{\frac{3}{1+\Dim}}\,.
\ee
(\ref{F8}) is a formula\footnote{It was common in the literature of that period to assign a single value to the fractal dimension introduced from another source, as in \cite{Kraichnan1985}.} that appears in Kraichnan \cite{Kraichnan1985}, but without any identification of the nature of $\mathbb{F}_{m}$. The size of the upper bound on $L\eta_{m}^{-1}$ ranges from the standard inverse Kolmogorov bound $Re^{3/4}$ when $m=1$ to $Re^{3}$ when $m = \infty$. 
\

\subsection{The two-dimensional case}

What of the two-dimensional case? Advantage can be taken of the lack of a vortex stretching term ($\bom\cdot\nabla\bu = 0$) in the vorticity formulation to find the enstrophy dissipation rate \cite{DF2002,Verma2019}
\bel{F9}
\nu L^{-2}\left<\int_{\mathbb{T}^{2}}|\nabla\bom|^{2}dV\right>_{T} \leq c\, \nu^{3}L^{-6}Re^{3}\,.
\ee
In a full $\mathbb{T}^{2}$ domain, (\ref{F9}) leads to an $Re^{1/2}$ bound on the set of inverse Kraichnan lengths $L\eta_{m,ens }^{-1}$. 
The equivalent of (\ref{F6}) for the enstrophy dissipation rate on $\mathbb{F}_{m}$ is
\bel{F10}
\varepsilon_{m,ens} = \nu L^{-\Dim}\left<\int_{\mathbb{F}_{m}}|\nabla\bom|^{2}dV\right>_{T}\,,
\ee
with $\Dim= 2/m$. Then $\varepsilon_{m,ens}\nu^{-3}$ can be expressed in terms of a set of inverse lengths $\eta_{m,ens}^{-1}$ such that $\varepsilon_{m,ens}\nu^{-3} = \eta_{m,ens}^{-4-\Dim}$, leading to
\bel{F11}
L\eta_{m,ens}^{-1} \leq c\,Re^{\frac{3}{4+\Dim}}\,,
\ee
and thus a spread of $Re^{1/2}$ to $Re^{3/4}$. In this two-dimensional case Alexakis and Doering \cite{AD2006} derived a tighter bound of $Re^2$ on the right hand side of (\ref{F9}) when the forcing has a single wave-number or has a constant flux. An amendation of the numerator in (\ref{F11}) gives a spread ranging from $Re^{1/3}$ to $Re^{1/2}$, thus implying even less multifractality. 

\section{\large Correspondence with the multifractal model}\label{MFM}

Kolmogorov's 1941 theory (K41) of three-dimensional, homogeneous, isotropic turbulence \cite{Frisch1995} revolves around the $p$th-order velocity structure function $\Sp$ defined as 
\bel{M1}
\Sp = \left<\left|\bu(\bx+\br) - \bu(\bx)\right|^{p}\right>_{stat.av.}\,.
\ee 
The scaling properties of $\Sp$ have their origin in the invariance property enjoyed by the incompressible Euler equations under the transformation $\bx' = \lambda^{-1}\bx$\,; $t' = \lambda^{h-1}t$ and $\bu = \lambda^{h}\bu'$. The scaling parameter $h$ can take any value, unlike the Navier-Stokes equations where it must satisfy $h=-1$. The invariance property shows that $\Sp$ scales as $\Sp \sim r^{hp}$. As explained in \cite{Frisch1995}, the axioms of K41 lead to $h=\third$, which is consistent with the ``four-fifths law'' $S_{3} = - \fourfifths\varepsilon r$. The problem with the result $\Sp \sim r^{p/3}$ is that experimental data deviate from the linear in $p$ scaling but instead lie on a concave curve $\zeta_{p}$ below the line $\third p$. The multifractal model of Parisi and Frisch \cite{PF1985,Frisch1995} adapted the K41 formalism by relaxing the requirement that $h=\third$ by insisting that the energy dissipation rate is invariant in $h$ only in an average sense. They achieved this by introducing the probability $P(h) \sim r^{3-D(h)}$ of observing a given scaling exponent $h$ at the scale $r$. Each value of $h$ belongs to a given fractal set of dimension $D(h)$. This procedure produces the result
\bel{M2}
\Sp \sim r^{\zeta_{p}}\quad\mbox{where}\quad
\zeta_{p} = \min_{h}\left[hp + 3 - D(h)\right]\,.
\ee
This must be constrained by the four-fifths law which requires $\zeta_{3} = 1$, thus leading to the inequalities for $D(h)$ and the co-dimension $C(h)$ (which sum to the spatial dimension)
\bel{M3}
D(h) \leq 3h + 2\quad\Rightarrow\quad C(h) \geq 1-3h\,.
\ee
Given that K41 and the MFM models revolve around the flow being homogeneous and isotropic, any comparison with results from the Navier-Stokes equations will likely only be valid when the flow has reached a fully turbulent state\,: i.e. when $T$ is sufficiently large. 
\par\smallskip
To make the two theories touch, it is natural to suggest a correspondence between $\Dim$ and $D(h)$ and $\Cim$ and $C(h)$, which leads to a simple inequality relating $m$ and $h$
\bel{D1a}
\frac{3}{m} \leq 2 + 3h\,,
\ee
from which we find $h \geq h_{min}$ with $h_{min} = m^{-1} - \twothirds$. The two opposite limits $m\to 1$ and $m\to \infty$ give
\bel{D2a}
-\twothirds \leq h_{min} \leq \third\,,
\ee
which is precisely the range found in \cite{DG2022}.
\par\smallskip
A more modern version of the multifractal model uses the theory of large deviations to re-express $P(h)$ as $P(h) \sim r^{C(h)}$, where $C(h)$ is the multifractal spectrum \cite{Frisch1995,Eyink2008,BD2019}. $C(h)$ therefore plays the role as the co-dimension. In the theory of large deviations it is possible that $C(h) \geq 3$, the domain dimension, corresponding to $D(h)<0$, but only with an infinitesimal probability. This would correspond to $m<0$, which must be excluded here. The exponent $3/(1+\Dim)$ in (\ref{F8}) is bounded below by
\bel{D3}
\frac{3}{1+\Dim} \geq \frac{1}{1+h}\,.
\ee
The left hand side of (\ref{D3}) is derived directly from the Navier-Stokes equations, while the right hand side comes from the MFM and can be recognized as the Paladin-Vulpiani inverse dissipation scale $\eta^{-1}_{PV}$ defined in (\ref{PVscale}). Paladin and Vulpiani \cite{PV1987a} treated (\ref{M3}) as an equality, which would change the $\geq$ in (\ref{D3}) to equality. Thus we see that the Paladin-Vulpiani scale appears naturally in this formulation.
\par\smallskip
Finally, given that there exists a continuum of sub-Kolmogorov scales from the $n=1$ case alone, the role of higher derivatives ($n >1 $) in the $X_{n,\Dim}$ in (\ref{F4}) is an open question and one that is relevant to the regularity problem. Dubrulle and Gibbon \cite{DG2022} have shown that the lower bound on $C(h) \geq 1-3h$ holds \textit{for all values of $n$}, meaning that one can do no better than the result given by the four-fifths law at $n=1$.  In (\ref{D2a}), $h \geq -\twothirds$ bounds $h$ away from the dangerous value of $h=-1$, where real singularities can occur \cite{CKN1982,Frisch1995,Eyink2008,DG2022}. Recent work in \cite{Ng2019,Ng2020} has characterized small structures using values of $h$. 
\par\bigskip\noindent
\textbf{Acknowledgments\,:} The author thanks the Isaac Newton Institute for Mathematical Sciences, Cambridge for support and hospitality during the programme \textit{Mathematical Aspects of Fluid Turbulence\,: where do we stand?} in 2022, when work on this paper was 
undertaken. This work was supported by EPSRC grant no EP/R014604/1\&\#34. Thanks are also due to Berengere Dubrulle (Saclay) and Dario Vincenzi (Nice) for discussions. I am indebted to Nadia Padhan and Kiran Kolluru of IISc Physics Department, Bangalore for the plots in Fig. 1.

\par\vspace{3mm}
\bibliographystyle{unsrt}

\end{document}